\documentclass[12pt]{article}
\usepackage{epsfig, amssymb}
\usepackage{graphicx,epsfig}
\begin{document}
\begin{titlepage}
\thispagestyle{empty}
\begin{flushright}
\end{flushright}\textbf{}

\bigskip

\begin{center}
\noindent{\Large \textbf
{Stochastic quantization and holographic Wilsonian renormalization group of scalar theories with arbitrary mass}}\\

\vspace{2cm} \noindent{
Jae-Hyuk Oh${}^{a}$\footnote{e-mail:jack.jaehyuk.oh@gmail.com}}

\vspace{1cm}
{\it
Department of Physics, Hanyang University \\
 Seoul 133-891, Korea${}^{a}$\\
}
\end{center}

\vspace{0.3cm}
\begin{abstract}
We have studied a mathematical relationship between holographic Wilsonian renormalization group(HWRG) and stochastic quantization(SQ) of scalar field with arbitrary mass in AdS spacetime. 
In the stochastic theory, the field is described by an equation with a form of harmonic oscillator with time dependent frequency and its Euclidean action also shows explicit time dependent  kernel in it.
We have obtained the stochastic 2-point correlation function and demonstrate that it reproduces the radial evolution of the double trace operator correctly via the suggested relation given in arXiv:1209.2242. Moreover, we justify our stochastic procedure with time dependent kernel by showing that it can be mapped to a new stochastic frame with a standard kernel without time dependence. Finally, we consider more general boundary conditions for the stochastic field to reproduce the radial evolution of the holographic boundary effective action when alternative quantization is allowed. We extensively study the Neumann boundary condition case and confirm that even in this case, the relation between HWRG and SQ is precisely hold. 
\end{abstract}
\end{titlepage}

\newpage

\tableofcontents

\section{Introduction}
\label{Introduction}
AdS/CFT correspondence has shed light on the various strongly coupled field theories by providing very useful insights on them.
Recently, Fluid/gravity duality and AdS/CMT have been widely studied and much useful information has been obtained. Among such studies, especially holographic Wilsonian renormalization group(HWRG)\cite{Heemskerk:2010hk,Faulkner:2010jy} provides the renormalization group flows of interesting multi-trace operators in the dual (conformal) field theories defined on a hypersurface with a certain radial cut-off in AdS space since it turns out that the radial cut-off in AdS space where the gravity theories are defined corresponds to the energy scale in the dual field theories. 

Another interesting challenge is to understand HWRG in the frame of stochastic quantization (SQ)\cite{Wu1,Paul1,Dijkgraaf:2009gr}, i.e. to figure out a mathematical relationship between them. The mathematical relation between SQ and HWRG has been addressed in several papers by J.Oh and D. P. Jatkar \cite{Oh:2012bx,Jatkar:2013uga,Oh:2013tsa} and
they developed the relation to the various holographic models in AdS space. The dictionary that the authors in \cite{Oh:2012bx} have found is that once we identify the boundary on-shell action, $I_{os}$(without holographic renormalization, i.e. by keeping divergent pieces in it) with Euclidean action, $S_E$ in SQ as
\begin{equation}
S_E=-2I_{os}, 
\end{equation}and request that the stochastic time $t$ is identified with $r$, the radial variable in AdS space as
\begin{equation}
r=t, 
\end{equation}
the stochastic procedure precisely reproduce the radial evolution of the double trace operators in the dual field theory defined on the $r=\epsilon$ hypersurface in AdS space.

In \cite{Oh:2012bx}, the authors provide two explicit examples to support their claim. One is massless scalar field in AdS$_2$ and another is one-form field in AdS$_4$. In these examples, they reproduced
the radial evolution of the double trace operators in the dual field theories from the stochastic 2-point correlation functions precisely via the following relationship:
\begin{equation}
\label{the RelaTion}
\langle \Psi_p(r)\Psi_{-p}(r)\rangle^{-1}_H=
\langle \Phi_p(t)\Phi_{-p}(t)\rangle^{-1}_S-\frac{1}{2}\frac{\delta^2 S_E}{\delta \Phi_p(t) \delta \Phi_{-p}(t)},
\end{equation}
where $(2\pi)^d\langle \Psi_p(r)\Psi_{-p}(r)\rangle^{-1}_H$ is the double trace operator ,$\langle \Phi_p(t)\Phi_{-p}(t)\rangle_S$ is the stochastic 2-point correlation function. 
The $\Psi$ is the field defined in AdS space and $\Phi$ is the stochastic field which should be identified by the relation given in \cite{Oh:2012bx}.

However, these examples are rather restrictive in a sense that their bulk action defined in AdS space effectively becomes theories of them defined on half of the flat space, $\mathbb R_+$
\footnote{The reason that it is not entire flat space is that the radial variable $r$ runs from 0 to $\infty$.}
once we substitute the explicit form of the AdS 
metric\footnote{
The form of the AdS$_{d+1}$ metric is that of Poincaré-patch, which is given by
\begin{equation}
ds^2=\frac{1}{r^2}(dr^2+\sum_{i=1}^{d}dx^idx^i).
\end{equation} 
}
into the action. In fact, these theories are the ones which are optimized to recover the relation. In the usual stochastic quantization there is no notion of metric which explicitly depends on the stochastic time $t$. The space of the stochastic quantization is a product space as $\mathbb M^d\times \mathbb R_+$, where $\mathbb M^d$ is $d$-dimensional manifold where the Euclidean action is defined and
$\mathbb R_+$ is a half of the real line for the stochastic time $t$. In the case of the two previous examples, the kinds of space of stochastic quantization are $\mathbb R\times \mathbb R_+$ for the massless scalar in AdS$_2$ and
 $\mathbb R^3\times \mathbb R_+$ for the one form field in AdS$_4$.

In \cite{Jatkar:2013uga}, the relation(\ref{the RelaTion}) is extended to conformally coupled scalar in AdS$_{d+1}$ space, which does not enjoy the properties that the previous examples present. 
One sickness in this example is that the Euclidean action shows explicit stochastic time dependence through the identification $S_E=-2I_{os}$. The formal form of the Euclidean action obtained by the identification is given by
\begin{equation}
S_E=\int g(t)\Phi^2(x)d^dx,
\end{equation}
where $\Phi$ is the stochastic field and $g(t)$ carries the time dependence in $S_E$. The form of Langevin equation is
\begin{equation}
\frac{\partial \Phi(x,t)}{\partial t}=-\frac{1}{2}\frac{\delta S_E}{\delta \Phi(x,t)}+\eta(x,t),
\end{equation}
where $\eta$ is the white Gaussian noise.

To evaluate the explicit form of the Langevin equation, we plug the Euclidean action into the Langevin equation and promote the field $\Phi(x)\rightarrow\Phi(x,t)$.
The Euclidean action before such promotion has no notion of stochastic time $t$ in it.
The purpose of the stochastic quantization is that one gets correlation functions in the very late time of $t$ as the consequence of the quantization of $S_E$. This means that the correlation function is that of $d$-dimensional theory. The information about the stochastic time $t$ is completely washed out by taking $t\rightarrow\infty$ in the correlators. Therefore, it may be non sense if there is $t$ dependence in the Euclidean action, $S_E$.

However,  one can avoid this sickness by using a field redefinition. This scalar theory is mapped to a theory of massless scalar field theory in $\mathbb R^d\times \mathbb R_+$ by an appropriate field redefinition. In this field frame, the Langevin equation and the Euclidean action show no explicit time dependence in them and the radial evolution of the double trace operator is precisely reproduced from the stochastic 2-point function via the relation(\ref{the RelaTion}). 

In this paper, we have extended this relation to scalar field theory in AdS$_{d+1}$ space with arbitrary mass. This theory shares the similar problem with the previous example. The Euclidean action obtained via the identification $S_E=-2I_{os}$ contains explicit time dependence. One can apply the same field redefinition used in \cite{Jatkar:2013uga} but one cannot remove the explicit time dependence in this case. The action with the redefined field has a form of harmonic oscillator with time dependent frequency due to the arbitrary mass term
\footnote{In conformally coupled scalar case, this time dependence is completely canceled out in the transformed field frame.}. This also implies that its boundary on-shell action has explicit cut-off dependence in it too(without holographic renormalization). 

A resolution for this problem is given in \cite{Haas}. The authors in \cite{Haas} suggest  a mapping from this frame to another of stochastic quantization, where there is no explicit time dependence on the Langevin equation and a new Gaussian noise satisfies the standard form of the correlation functions, for example, $\langle \eta_p(t)\eta_{p^\prime}(t^\prime) \rangle=\delta(t-t^\prime)\delta^d(p-p^\prime)$. 
Such map is achieved by a rescaling of the stochastic time as well as  appropriate field redefinition. This idea applies to the scalar field theory with arbitrary mass and we have shown an example of the explicit mapping in the last section in this note. The method justifies that one can construct Langevin equation out of the Euclidean action with time dependent kernel in it. By using such time dependent Euclidean action from the relation $S_E=-2I_{os}$, we have reproduced the radial evolution of the double trace operator from stochastic process precisely.

One special property of the scalar theory with arbitrary mass in AdS is that the solution of the Langevin equation has no more exponential form of the function. Usually, the Langevin equation is 
a type of diffusion equation and its solution is a form of $e^{-t}$. However, in the case of the scalar field with arbitrary mass, the on-shell action provides an Euclidean action where its kernel contains a combination of Bessel functions and their derivatives. Such nontrivial Euclidean action gives the correctly reproduce the holographic boundary effective action.

For the final issue, we consider more general boundary conditions for the stochastic fields. The motivation of this is that when the scalar field mass is in the range of $-\frac{d^2}{4}\leq m^2 \leq-\frac{d^2}{4}+1$ in AdS$_{d+1}$ space, one can impose more diverse boundary conditions for the field in the dual gravity in holographic framework because alternative quantization is possible. In the literatures so far\cite{Oh:2012bx,Jatkar:2013uga,Oh:2013tsa}, the case of standard quantization is studied only, where the authors consider only Dirichlet boundary condition at the initial stochastic time, $\phi(t=t_0)=0$. Among many of the boundary conditions, we concentrate on Neumann boundary condition and evaluate the radial evolution of the boundary effective action by using the suggested relation as
\begin{equation}
\label{brief-main}
S_B= \int^t_{t_0} d\tilde t \int d^dx \mathcal L _{FP}(\phi(\tilde t,x)),
\end{equation}
where $\mathcal L_{FP}$ is Fokker-Planck Lagrangian. When the Dirichlet boundary condition($\phi(t=t_0)=0$) is imposed for the stochastic field, the double trace part of the boundary effective action is captured only. It turns out that the boundary condition, $\phi(t=t_0)=0$ is very specific, which kills the single trace part in the boundary effective action. More relieved Dirichlet boundary condition as $\phi(t=t_0)=\phi_0$ will not kill the single trace part.
Once the Neumann boundary condition is requested, one can evaluate the single trace part too. It turns out that for the case that the Neumann boundary condition is imposed, the double and single trace parts of the operators in the boundary effective action are correctly reproduced by the stochastic prescription.

In Sec.\ref{Holographic Wilsonian renormalization group}, we review HWRG for the scalar field with arbitrary mass in AdS$_{d+1}$ and the radial evolution of its boundary effective action. In Sec.\ref{Stochastic Quantization}, we develop stochastic quantization for the theory and show that the  stochastic 2-point correlation function reproduces the radial evolution of the double trace operator correctly. In Sec.\ref{Stochastic qunatization with time dependent kernel}, we discuss a justification for the stochastic process performing in Sec.\ref{Stochastic Quantization} with time dependent kernel by showing that the primitive Langevin equation is mapped to a new type of that without explicit time dependence and the standard form of correlations with the transformed white Gaussian noise. Finally, in Sec.\ref{Alternative quantization and the related stochastic business}, we discuss an application of Neumann boundary condition for the stochastic field when alternative quantization is possible in the dual field theories in the holographic frame work.

\section{Holographic Wilsonian renormalization group}
\label{Holographic Wilsonian renormalization group}
We start with a free massive scalar field action defined in AdS$_{d+1}$ as
\begin{equation}
S=\int_{r>\epsilon}drd^dx\sqrt{g}\mathcal L(\phi,\partial \phi) +S_B,
\end{equation}
where $r$ is AdS radial coordinate, $\epsilon$ is radial cut-off and $S_B$ is the boundary effective action at $r=\epsilon$.
$\mathcal L$ is the Lagrangian density of the
massive scalar field defined in the AdS space, which is given by
\begin{equation}
\mathcal{L} =\frac{1}{2} g^{\mu\nu} \partial_\mu\phi\partial_\nu\phi
+\frac{1}{2}m^2\phi^2,
\end{equation}
where $g_{\mu\nu}$ is the AdS metric:
\begin{equation}
\label{AdS-metric}
ds^2=g_{\mu\nu}dx^\mu dx^\nu=\frac{dr^2+\sum^d_{i=1}dx^idx^i}{r^{2}}.
\end{equation}
The bulk spacetime indices $\mu$, $\nu$ run from 1 to $d+1$, we define that $x^{d+1}\equiv r$ and  $x^1$
... $x^{d}$ are boundary directional coordinates. 

\paragraph{Flow of the effective action in dual CFT\cite{Heemskerk:2010hk,Faulkner:2010jy}}

The boundary effective action is sum of boundary multi trace operators multiplied by the boundary value of the bulk field $\phi$ with the boundary momentum integration at $r=\epsilon$ hypersurface. Therefore, 
\begin{equation}
\label{SB-ansatz}
S_B=\Lambda(\epsilon)+\int\frac{d^dp}{(2\pi)^d}J(\epsilon,p)\phi_{-p}(\epsilon)-\frac{1}{2}\int \frac{d^dp}{(2\pi)^d} D(p,\epsilon)\phi_p(\epsilon)\phi_{-p}(\epsilon)+{\rm \ multi\ trace\ part},
\end{equation} 
where $J(p,\epsilon)$, $D(p,\epsilon)$ are the single and double trace operators respectively.
For the free field in the bulk, we have at most double trace operators and we are interested in the single and double trace parts of the boundary effective action
\footnote{The term $\Lambda(\epsilon)$ is called boundary cosmological constant, which is the field independent.}. 
The Hamilton-Jacobi equation for the boundary effective action is given by
\begin{equation}
\label{hJCO-eQ}
\partial_\epsilon S_B=-\int_{r=\epsilon}d^dp \left( \frac{1}{2\sqrt{g}g^{rr}}\left(\frac{\delta S_B}{\delta \phi_p}\right)\left(\frac{\delta S_B}{\delta \phi_{-p}}\right)-\frac{1}{2}\sqrt{g}(g^{ij}p_ip_j+m^2)\phi_p\phi_{-p}\right).
\end{equation}
By substitution of (\ref{SB-ansatz}) into (\ref{hJCO-eQ}), we get
\begin{eqnarray}
\partial_\epsilon \Lambda(\epsilon)&=&-\frac{1}{2}\int \frac{d^dp}{(2\pi)^{2d}}\frac{1}{\sqrt{g}g^{rr}}J(\epsilon,-p)J(\epsilon,p), \\ 
\partial_\epsilon J(\epsilon,p)&=&\frac{1}{\sqrt{g}g^{rr}(2\pi)^d}J(\epsilon,p)D(\epsilon,p),\\
\partial_\epsilon D(p,\epsilon)&=&\frac{1}{\sqrt{g}g^{rr}(2\pi)^d}D(\epsilon,p)D(\epsilon,-p)-(2\pi)^d\sqrt{g}(r^2p^2+m^2)
\end{eqnarray}
and their solutions are given by
\begin{eqnarray}
\Lambda(\epsilon)&=&-\frac{1}{2}\int^\epsilon d\epsilon^\prime\int \frac{d^dp}{(2\pi)^{2d}}\frac{1}{\sqrt{g}g^{rr}}\frac{\beta_p\beta_{-p}}{\phi_p\phi_{-p}} , \\
J(\epsilon,p)&=&- \frac{\beta_p}{\phi_p} \\
D(\epsilon,p)&=&-(2\pi)^d\frac{\Pi_\phi}{\phi_p},
\end{eqnarray}
where $\beta_p$ is an arbitrary constant and $\Pi_\phi$ is the canonical momentum of the field $\phi$. The canonical momentum satisfies following equations:
\begin{equation}
\Pi_\phi=\sqrt{g}g^{rr}\partial_r\phi {\rm \ \ and\ \
} \partial_r\Pi_\phi=\sqrt{g}(r^2p^2+m^2)\phi,
\end{equation}
where the first equation is nothing but definition of the canonical momentum. By combining these two equations,
one can reconstitute the bulk equation of motion as
\begin{equation}
\partial_r(r^{1-d}\partial_r\phi)=\frac{1}{r^{d+1}}(r^2p^2+m^2)\phi,
\end{equation}
as well.

The solution of this equation in the zero boundary momentum $p=0$ case is
\begin{equation}
\phi(r)=b_1 r^{\frac{d}{2}-\nu}+b_2  r^{\frac{d}{2}+\nu},
\end{equation} 
where $b_1$ and $b_2$ are arbitrary constants and
\begin{equation}
\nu\equiv\frac{1}{2}\sqrt{d^2+4m^2}.
\end{equation}
Once we turn on the boundary directional momentum $p$ of the field, then the solution is a linear combination of the Bessel $K$ and $I$ as
\begin{equation}
\phi_p(r)=c_1r^{d/2}K_{\nu}(|p|r)+c_2r^{d/2}I_{\nu}(|p|r),
\end{equation}
where $c_1$ and $c_2$ are $p$ dependent constants.


\paragraph{Field redefinition}
we discuss HWRG in different field frame. In this field frame, the massive scalar field in AdS space becomes harmonic oscillator with $r$ dependent frequency.
By using the explicit form of the metric(\ref{AdS-metric}), the bulk action is written as
\begin{equation}
S=\frac{1}{2}\int drd^dp \left(r^{1-d}\partial_r \phi_p \partial_r \phi_{-p}+p^2r^{1-d}\phi_p\phi_{-p} +m^2 r^{-1-d}\phi_p\phi_{-p}\right).
\end{equation}
We define a new field $f_p$ as $\phi_p=r^{\frac{d-1}{2}}f_p$, and substitute it into the action. Then the action is transformed to
\begin{equation}
\label{bulk action with f}
S=\frac{1}{2}\int dr d^dp \left[{\partial_r f_p}{\partial_r f_{-p}}+p^2f_pf_{-p}+\frac{1}{r^2}(m^2+\frac{d^2-1}{4})f_pf_{-p}\right]
\end{equation}
Its equation of motion is given by
\begin{equation}
\label{redefined field equaiton}
0=-\partial_r\partial_r f_{p}+p^2f_p+r^{-2}\left(m^2+\frac{d^2-1}{4}\right)f_p,
\end{equation}
which is a form of that of harmonic oscillator with $r$-dependent frequency $\omega(r)$, where $\omega^2(r)=p^2+r^{-2}(m^2+\frac{d^2-1}{4})$.
The solutions of this equation are
\begin{eqnarray}
\label{solution of redefined field f p=o}
f(r)&=&b_1r^{\frac{1}{2}-\nu}+b_2r^{\frac{1}{2}+\nu}{\rm \ \ for\ zero\ momentum\ case}, \\ 
\label{solution of redefined field f}
f_p(r)&=&c_1r^{1/2}K_{\nu}(|p|r)+c_2r^{1/2}I_{\nu}(|p|r)
{\rm\ for\ }p\neq0,
\end{eqnarray}
where $b_1$ and $b_2$ are arbitrary constants and $c_1$ and $c_2$ are arbitrary boundary momentum $p$ dependent functions.

HWRG in this field frame is given as follows. 
We prepare a ansatz for the boundary effective action in this field frame as
\begin{equation}
\label{SB-ansatzf}
S^\prime_B=\Lambda_f(\epsilon)+\int\frac{d^dp}{(2\pi)^d}J_f(\epsilon,p)f_{-p}(\epsilon)-\frac{1}{2}\int \frac{d^dp}{(2\pi)^d} D_f(p,\epsilon)f_p(\epsilon)f_{-p}(\epsilon)+{\rm \ multi\ trace\ part},
\end{equation}
The Hamilton Jacobi equation in this field frame is given by
\begin{equation}
\partial_\epsilon S^\prime_B=-\frac{1}{2}\int_{r=\epsilon}d^dp\left( \left(\frac{\delta S^\prime_B}{\delta f_p}\right) \left(\frac{\delta S^\prime_B}{\delta f_{-p}}\right)-\left(|p|^2+\frac{1}{r^2}\left(m^2+\frac{d^2-1}{4}\right)\right)f_p f_{-p}\right).
\end{equation}
By plugging this into the action(\ref{bulk action with f}), we get
equations of motion of the single and double trace operators, $J_f$ and $D_f$ as
\begin{eqnarray}
\partial_\epsilon J_f(p,\epsilon)&=&\frac{1}{(2\pi)^d}J_f(p,\epsilon) D_f(p,\epsilon), \\
\partial_\epsilon D_f(p,\epsilon)&=& \frac{1}{(2\pi)^d}D_f(p,\epsilon)D_f(-p,\epsilon)-(2\pi)^d\left[p^2+\frac{1}{r^2}\left(m^2+\frac{d^2-1}{4}\right)\right],
\end{eqnarray}
where $\Pi_f$ is canonical momentum of the field $f_p$, which satisfies the following equations:
\begin{equation}
\label{canonical moemntaof f}
\Pi_f=\partial_r f_p {\rm\ \ and\ \ }\partial_r \Pi_f=p^2f_p+\frac{1}{r^2}\left(m^2+\frac{d^2-1}{4}\right).
\end{equation}

In the zero boundary momentum case, 
the solution of the single and double trace operators are given by
\begin{eqnarray}
\label{double trace solution with p=0}
J_f(r)&=& -\frac{\beta_p}{f_p}=-\frac{\beta_p}{d_1 r^{\frac{1}{2}-\nu}+d_2 r^{\frac{1}{2}+\nu}},\\
D_f(r)&=&-(2\pi)^d \frac{\partial_r f(r)}{f(r)}=-(2\pi)^d\frac{d_1(\frac{1}{2}-\nu)+d_2(\frac{1}{2}+\nu)r^{2\nu}}{d_1 r+d_2r^{2\nu+1}},
\end{eqnarray}
where $d_1$ and $d_2$ and $\beta_p$ are arbitrary constants.
Then, the double trace part of the boundary effective action is given by
\begin{equation}
\label{boundary eff action for 0 momentum}
S_B=\frac{1}{2}\frac{d_1(\frac{1}{2}-\nu)+d_2(\frac{1}{2}+\nu)r^{2\nu}}{d_1 r+d_2r^{2\nu+1}}f^2
\end{equation}

Once we turn on the boundary directional momentum $p$, the solution of the single and double trace operators are given by
\begin{eqnarray}
\label{double-trace radial evolution non-zero Momentum}
J(p,\epsilon)&=& -\frac{\beta_p}{f_p}=-\frac{\beta_p}{\epsilon^{1/2}(c_1 K_{\nu}(|p|\epsilon)+c_2I_{\nu}(|p|\epsilon))} \\
D_f(p,\epsilon)&=&-(2\pi)^d\frac{\Pi_f}{f_p}=-(2\pi)^d\partial_\epsilon\ln
[\epsilon^{1/2}(c_1 K_{\nu}(|p|\epsilon)+c_2I_{\nu}(|p|\epsilon))].
\end{eqnarray}
and so the double trace part of the boundary effective action becomes
\begin{equation}
\label{beztion}
S_B=-\frac{1}{2(2\pi)^d}\int d^dp D_f(p,\epsilon)f^{(0)}_p(\epsilon)f^{(0)}_{-p}(\epsilon).
\end{equation}


\paragraph{On-shell action} We discuss zero momentum case first. To evaluate the on-shell action,  we need to choose regular solution(which should not be divergent in AdS interior). Therefore, $b_2=0$ is chosen in (\ref{solution of redefined field f p=o}). Then, the form of the regular solution is given by
\begin{equation}
f(r)=f^{(0)}\frac{r^{\frac{1}{2}-\nu}}{\epsilon^{\frac{1}
{2}-\nu}},
\end{equation}
where $f^{(0)}$ is boundary value of the bulk field $f(r)$ at $r=\epsilon$.
The on-shell action is, by definition, the bulk action upto equation of motion evaluated on $r=\epsilon$ hypersurface, which is given by
\begin{equation}
\label{on-shell ACTION}
I_{os}=\frac{1}{2}\int_{r=\epsilon}d^dpf_p \partial_r f_{-p}.
\end{equation}
We plug the regular bulk solution into this and we get
\begin{equation}
\label{ON-Shell p=0}
I_{os}= \frac{1}{2\epsilon}\left(\frac{1}{2}-\nu\right)\left(f^{(0)}\right)^2
\end{equation}

For nonzero momentum ($p\neq0$) case, we set $c_2=0$ for the bulk solution to be regular. We rewrite this regular solution in terms of boundary value of the field $\phi$ as
\begin{equation}
f_p(r)=f^{(0)}_p\frac{r^{1/2}K_{\nu}(|p|r)}{\epsilon^{1/2}K_{\nu}(|p|\epsilon)},
\end{equation}
where $f^{(0)}_p$ is the boundary value of the bulk field $f_p(r)$ at $r=\epsilon$.
By substituting this regular solution into the formal form of the on-shell action(\ref{on-shell ACTION}), the on-shell action is given by
\begin{equation}
\label{non-zero momentum on-shell ACTION}
I_{os}=\frac{1}{2}\int_{r=\epsilon} d^dp
f^{(0)}_pf^{(0)}_{-p}\partial_r\ln[r^{1/2} K_{\nu}(|p|r)].
\end{equation}

\section{Stochastic quantization}
\label{Stochastic Quantization}
In this section, we develop stochastic quantization by identifying the Euclidean action, $S_E$ with the on-shell actions (\ref{ON-Shell p=0}) or (\ref{non-zero momentum on-shell ACTION}) through the suggested relation $S_E=-2I_{os}$. It will be shown that the radial evolution of the double trace operator can be
reproduced by stochastic 2-point correlation function via the relation suggested in \cite{Oh:2012bx}
\footnote{The relation is given by the equation (2.38) in \cite{Oh:2012bx}.}. It is given by
\begin{equation}
\label{the relation}
\langle f^{(0)}_p(r)f^{(0)}_{-p}(r)\rangle^{-1}_H=
\langle \Phi^{(0)}_p(t)\Phi^{(0)}_{-p}(t)\rangle^{-1}_S-\frac{1}{2}\frac{\delta^2 S_E}{\delta \Phi_p(t) \delta \Phi_{-p}(t)},
\end{equation}
provided by $r=t$ identification. $\langle \Phi^{(0)}_p(t)\Phi^{(0)}_{-p}(t)\rangle_S$ is the stochastic 2-point correlation function and $\langle f^{(0)}_p(r)f^{(0)}_{-p}(r)\rangle_H=\frac{\delta^2 S_B}{\delta f^{(0)}_p \delta f^{(0)}_{-p}}$, which is proportional to the double trace operator.

\subsection{Stochastic partition function}
We start with the Langevin equation, which is given by
\begin{equation}
\label{Langevin-equation}
\frac{\partial \Phi(x,t)}{\partial t}=-\frac{1}{2}\frac{\delta S_E}{\delta \Phi(x,t)}+\eta(x,t),
\end{equation}
where $t$ is the stochastic time, $\Phi$ is the stochastic field, $\eta$ is the white Gaussian noise and $S_E$ is the Euclidean action. We
take this Euclidean action to be the following form: 
\begin{equation}
\label{Euclidean action}
S_E=
\int d^dx\ g(t)\Phi^2(x,t),
\end{equation}
which contains stochastic time dependent kernel $g(t)$ in it. By using such a form of the Euclidean action, the Langevin equation
becomes
\begin{equation}
\label{langevin equation with g}
\frac{\partial \Phi(x,t)}{\partial t}=-g(t)\Phi(x,t)+\eta(x,t).
\end{equation}
To evaluate the stochastic correlation functions correctly, we consider the stochastic partition function.
The stochastic partition function is
\begin{equation}
Z[\eta]=\int \mathcal D[\eta]\exp\left(- \mathcal S \right),
\end{equation}
where
\begin{equation}
\mathcal S=\frac{1}{2}\int^t_{t_0} dtd^dx\eta^2(x,t),
\end{equation}
and $t$ and $t_0$ are the final and initial stochastic time for the stochastic process respectively. From this partition function, the stochastic 2-point correlation function is given by
\begin{equation}
\langle \eta_p(t)\eta_{p^\prime}(t^\prime) \rangle=\frac{1}{Z}\int D[\eta]\eta_p(t)\eta_{p^\prime}(t^\prime)e^{-\mathcal S}=\delta(t-t^\prime)\delta^d(p-p^\prime).
\end{equation}
By using the Langevin equation, we replace the stochastic noise, $\eta$ with the stochastic field, $\Phi$ and then we have
\begin{eqnarray}
\mathcal S&=&\int^t_{t_0} dtd^dx \left( \frac{1}{2}(\partial_t\Phi)^2 +\frac{1}{2}(g^2(t)-\partial_t g(t))\Phi^2 +\frac{1}{2}\partial_t(g(t)\Phi^2)\right),\\ \nonumber
&\equiv& \left. S_{FP}+S_E\right|^{t=t}_{t=t_0},
\end{eqnarray}
where $S_{FP}$ is called the Fokker-Planck action which is given by
\begin{equation}
S_{FP}=\int dtd^dx \mathcal L_{FP}
\end{equation}
and $\mathcal L_{FP}$ is the Fokker-Planck Lagrangian density:
\begin{equation}
\mathcal L_{FP}= \frac{1}{2}(\partial_t\Phi)^2 +\frac{1}{2}(g^2(t)-\partial_t g(t))\Phi^2.
\end{equation}
One can define this Fokker-Planck action in the momentum space too, by
using the Fourier transform as 
\begin{equation}
\Phi(x,t)=\frac{1}{(2\pi)^{d/2}}\int d^dpe^{-ipx}\Phi_p(t),
\end{equation}
then the Fokker-Planck action becomes
\begin{equation}
\label{Fokker-Planck action}
S_{FP}=\frac{1}{2}\int dtd^dp[\partial_t \Phi_p \partial_t \Phi_{-p}+(g^2(t)-\partial_t g(t))\Phi_p\Phi_{-p}]
\end{equation}

According to \cite{Oh:2012bx}, this Fokker-Planck action is identified with the
action(\ref{bulk action with f}). For this, we demand that the Ricatti term
appearing in (\ref{Fokker-Planck action}) is equal to the radial
coordinate dependent frequency term in (\ref{bulk action with f}) along
with the identification of the AdS radial coordinate $r$ with the
stochastic time coordinate $t$.  This is a part of the SQ and AdS/CFT dictionary.  In other words, we replace the
radial coordinate $r$ by the stochastic time $t$. This gives us the
equation,
\begin{equation}
\label{g-equation}
g^2(t)-\partial_t g(t)=p^2+\frac{1}{t^2}\left( m^2+\frac{d^2-1}{4} \right).
\end{equation}
For the zero boundary momentum case, $p=0$ the solution of $g(t)$ is either
\begin{equation}
g_1(t)= \frac{-\frac{1}{2}+\nu}{t}, {\rm \ \ or \ \ }g_2(t)=\frac{-\frac{1}{2}-\nu}{t}.
\end{equation}
Once we choose the solution $g_1(t)$, then the Euclidean action(\ref{Euclidean action}) is consistent with the prescription for the choice of the Euclidean action 
from the on-shell action given in \cite{Oh:2012bx}. The choice of the solution $g_2(t)$ corresponds to turning on the irregular part of the scalar field solution in AdS space, meaning that it causes divergence at the AdS center, $r=\infty$.
The boundary value of the bulk scalar field $f_(r)$ is identified to be the stochastic field. 
For the nonzero momentum case, the solution of this equation
is
\begin{equation}
\label{g-solution}
g(t)=-\partial_t\log[\sqrt{t}\ K_{\nu}(|p|t)+a_0
\sqrt{t}\ I_{\nu}(|p|t)]\ .
\end{equation}
To match this with the on-shell action (\ref{non-zero momentum on-shell ACTION}), we take $a_0$ to vanish. In fact, the term being proportional to $a_0$ comes from the irregular part of the bulk solution in the on-shell action. Therefore, a choice of $a_0=0$ is consistent with the Euclidean action derived from the on-shell action.

\subsection{The stochastic 2-point correlation function and the boundary effective action}
\label{The stochastic 2-point correlation function and the boundary effective action}
In this subsection, we compute the stochastic 2-point correlation function to match with the boundary effective action in HWRG computation.
\paragraph{Langevin approach} We discuss the zero boundary momentum case first.
By substitution of the Euclidean action,
\begin{equation}
S_E=-\frac{1}{t}\left(\frac{1}{2}-\nu \right)(f^{(0)})^2
\end{equation}
obtained from the on-shell action(\ref{ON-Shell p=0}) and by applying the identification $r=t$, the Langevin equation(\ref{Langevin-equation}) is given by
\begin{equation}
\label{langevin equation with p=0}
\frac{\partial \Phi(t)}{\partial t}=\frac{1}{t}(1-2\nu)\Phi(t)+\eta(t),
\end{equation}
and its solution is
\begin{equation}
\Phi(t)=\int^t_{t_0}\frac{t^{\frac{1}{2}-\nu}}{t^{\prime \frac{1}{2}-\nu}}\eta(t^\prime)dt^\prime,
\end{equation}
where $t_0$ is initial time, which will determined by a initial boundary condition soon. Now, we compute stochastic 2-point correlation function as
\begin{eqnarray}
\langle \Phi(t)\Phi(t^\prime)\rangle_S&=&\int^t_{t_0}\int^{t^\prime}_{t^\prime_0}\frac{t^{\frac{1}{2}-\nu}t^{\prime\frac{1}{2}-\nu}}{t^{\frac{1}{2}-\nu}_0t^{\prime\frac{1}{2}-\nu}_0}\langle \eta(t_0)\eta(t^\prime_0)\rangle dt_0 dt^\prime_0 \\ \nonumber
&=&\int^t_{t_0}\frac{t^{\frac{1}{2}-\nu}t^{\prime\frac{1}{2}-\nu}}{t^{1-2\nu}_0}dt_0,\\ \nonumber
&=&\frac{t^{\frac{1}{2}-\nu}t^{\prime\frac{1}{2}-\nu}}{2\nu}(t^{2\nu}-t^{2\nu}_0).
\end{eqnarray}
We have used the relation of 2-point function of the white Gaussian noise $\eta$ for the second equality in the above computation:
\begin{equation}
\label{correlation with p=0}
\langle\eta(t)\eta(t^\prime) \rangle=\delta(t-t^\prime).
\end{equation}
Then, the equal time commutator is
\begin{equation}
\langle \Phi(t)\Phi(t) \rangle_S=\frac{t^{1-2\nu}}{2\nu}(t^{2\nu}-t^{2\nu}_0).
\end{equation}

It is manifest that this stochastic 2-point correlation function can reproduce the solution of double trace operator(\ref{double trace solution with p=0}) via the suggested relation in \cite{Oh:2012bx} as
\begin{equation}
\langle f^{(0)}(r)f^{(0)}(r)\rangle^{-1}_H=\langle \Phi(t)\Phi(t)\rangle^{-1}_S-\frac{1}{2}\frac{\delta^2 S_E}{\delta \Phi(t) \delta \Phi(t)},
\end{equation} 
provided by the identification $r=t$ and requesting a initial boundary condition for $t_0$ as $t_0=\left(-\frac{d_1}{d_2}\right)^{\frac{1}{2\nu}}$.

Next, we discuss nonzero momentum case. We start with the Euclidean action obtained from the on-shell action (\ref{non-zero momentum on-shell ACTION}) via the identification $S_E=-2I_{os}$. From this form of the Euclidean action, we derive the Langevin equation as
\begin{equation}
\frac{\partial \Phi_p(t)}{\partial t}=\partial_t \ln(t^{1/2}K_{\nu}(|p|t))\Phi_p(t)+\eta_p(t).
\end{equation}
Solution of this equation is
\begin{equation}
\phi_p(t)=\int^t_{t_0}\frac{t^{1/2}K_\nu(|p|t)}{t^{\prime1/2}K_\nu(|p|t^\prime)}\eta_p(t^\prime)dt^\prime.
\end{equation}
Let us compute stochastic 2-point correlation function, which is given by
\begin{eqnarray}
\langle \phi_p(t)\phi_{p^\prime}(t^\prime) \rangle&=&\int^t_{t_0}\int^{t^\prime}_{t_0}\frac{t^{1/2}t^{\prime1/2}K_\nu(|p|t)K_\nu(|p|t^\prime)}{\bar t^{1/2}\bar t^{\prime1/2}K_\nu(|p|\bar t)K_\nu(|p|\bar t^\prime)}\langle \eta_p(\bar t) \eta_{p^\prime}(\bar t^\prime)\rangle d\bar td\bar t^\prime \\ \nonumber
&=&\int^t_{t_0}\frac{t^{1/2}t^{\prime1/2}}{\bar t}\frac{K_\nu(|p|t)K_\nu(|p|t^\prime)}{K_\nu(|p|\bar t)^2}d\bar t{\ }\delta^d(p-p^\prime)\\ \nonumber
&=&t^{1/2}t^{\prime1/2}K_\nu(|p|t)K_\nu(|p|t^\prime)\left.\frac{I_\nu(|p|\bar t)}{K_\nu(|p|\bar t)}\right|^{\bar t=t}_{\bar t=t_0},
\end{eqnarray}
where for the second equality, we have used the noise correlator,
\begin{equation}
\label{etaeta}
\langle \eta_p(t)\eta_{p^\prime}(t^\prime)\rangle=\delta(t-t^\prime)\delta^d(p-p^\prime).
\end{equation}
For the third equality, the following integral formula has been used:
\begin{equation}
\int \frac{dx}{x[K_\alpha(x)]^2}=\frac{I_\alpha(x)}{K_\alpha(x)}.
\end{equation}
We are interested in the equal time commutator, so we take $t^\prime=t$. The equal time commutator is given by
\begin{equation}
\label{st-2-pint func non-zero momentum}
\langle \phi_p(t)\phi_{p^\prime}(t) \rangle= \delta^d(p-p^\prime){\ }t[K_\nu(|p|t)I_\nu(|p|t)-\beta(t_0)K_\nu(|p|t)^2],
\end{equation}
where 
\begin{equation}
\beta(t_0)=\frac{I_\alpha(|p|t_0)}{K_\alpha(|p|t_0)}.
\end{equation}

To check if the stochastic correlation function(\ref{st-2-pint func non-zero momentum}) reproduces the radial evolution of the double trace operator(\ref{double-trace radial evolution non-zero Momentum}) correctly via the relation(\ref{the relation}), we compute
\begin{eqnarray}
\label{computation of L - Se}
\langle \Phi_p(t) \Phi_{-p}(t) \rangle-\frac{1}{2}\frac{\delta^2 S_E}{\delta \Phi_p(t) \delta \Phi_{-p}(t)}=
\partial_t \ln [t^{1/2}(-\beta(t_0)K_\nu(|p|t)+I_\nu(|p|t))].
\end{eqnarray}

Once we identify the radial variable $r$ with the stochastic time $t$ and request a initial boundary condition, $\beta(t_0)=-\frac{c_1}{c_2}$,
then (\ref{computation of L - Se}) perfectly match with (\ref{double-trace radial evolution non-zero Momentum}).

\paragraph{Fokker-Planck approach} 
The Fokker-Planck action is
given by
\begin{equation}
\label{FP-action-nu}
S_{FP}=\frac{1}{2}\int dtd^dp\left[\partial_t \Phi_p \partial_t \Phi_{-p}+\left(|p|^2+\frac{\nu^2-\frac{1}{4}}{t^2}\right)\Phi_p\Phi_{-p}\right],
\end{equation} 
and it equation of motion is 
\begin{equation}
\label{eq-of-mo-nu}
0=-\partial^2_t\Phi_p(t)+\left[| p|^2 + \frac{\nu^2-\frac{1}{4}}{t^2}\right]\Phi_p(t).
\end{equation}
By using its equation of motion, we get the on-shell evaluation of this action, which is
\begin{equation}
S_{FP}=\frac{1}{2}\left.\int d^dp{\ } \Phi_p(t^\prime)\partial_{t^\prime}\Phi_{-p}(t^\prime)\right|^{t^\prime=t}_{t^\prime=t_0},
\end{equation}
where $\Phi(t)$ becomes the most general solution of the equation of motion. 
For the zero boundary momentum case, the most general solution of $\Phi(t)$ is given by
\begin{equation}
\label{zero momen sto sol}
\Phi(t^\prime)=\Phi(t)\frac{h_1t^{\prime\frac{1}{2}-\nu}+h_2t^{\prime\frac{1}{2}+\nu}}{h_1t^{\frac{1}{2}-\nu}+h_2t^{\frac{1}{2}+\nu}},
\end{equation}
where $h_1$ and $h_2$ are arbitrary constants.
This solution is designed to satisfy the stochastic boundary condition at $t^\prime=t$ as $\Phi(t^\prime=t)=\Phi(t)$. To reproduce the double trace part of the boundary effective action in 
HWRG, we have chosen the initial boundary condition too as
\begin{equation}
t_0=\left(-\frac{h_1}{h_2}\right)^{\frac{1}{2\nu}}.
\end{equation}
Once the solution(\ref{zero momen sto sol}) is plugged into the Fokker-Planck action, we get
\begin{equation}
S_{FP}=\frac{1}{2}\frac{h_1\left(\frac{1}{2}-\nu\right)+h_2\left(\frac{1}{2}+\nu\right)t^{2\nu}}{h_1t+h_2t^{2\nu+1}}.
\end{equation}
Once we identify $d_1=h_1$ and $d_2=h_2$, the this Fokker-Planck action correctly reproduce the 
boundary effective action, $S_B$ given in (\ref{boundary eff action for 0 momentum}) provided by $r=t$.

Next we discuss nonzero momentum case ($p\neq0$).
\ref{FP-action-nu},
From the equation of motion(\ref{eq-of-mo-nu}), the most general solution for this case is obtained as 
\begin{equation}
\Phi_p(t)=d_1t^{1/2}K_{\nu}(|p|t)+d_2t^{1/2}I_{\nu}(|p|t),
\end{equation}
where $d_1$ and $d_2$ are $p$-dependent arbitrary constants. For this solution, we impose stochastic boundary condition as
$\Phi(\tilde t=t)=\Phi(t)$, then we have
\begin{equation}
\Phi_p(\tilde t)=\Phi(t)\frac{\tilde
  t^{1/2}[d_1K_{\nu}(|p|\tilde t)+d_2I_{\nu}(|p|\tilde
  t)]}{t^{1/2}[d_1K_{\nu}(|p|t)+d_2
  I_{\nu}(|p|t)]}.
\end{equation}
By substituting this solution into the Fokker-Planck action, we obtain
\begin{equation}
\label{foker-planc}
S_{FP}=\frac{1}{2}\int d^dp\frac{\partial_t[t^{1/2}(d_1K_{\nu}(|p|t)+d_2
  I_{\nu}(|p|t))]}{t^{1/2}(d_1K_{\nu}(|p|t)+d_2
  I_{\nu}(|p|t))}
\end{equation}
by requesting a initial boundary condition as
\begin{equation}
\label{initial-ki}
\frac{d_1}{d_2}=-\frac{K_\nu(t_0)}{I_\nu(t_0)}.
\end{equation}
In sum, once we take $d_1=c_1$, $d_2=c_2$ and $r=t$ then (\ref{foker-planc}) reproduces (\ref{beztion}) precisely.

\section{Stochastic quantization with time dependent kernel}
\label{Stochastic qunatization with time dependent kernel}
In this section, we justify our computations in the previous sections by providing 
a resolution for the subtle issue that we have mentioned in introduction: the issue of the explicit time dependent kernel in the Euclidean action. 

\paragraph{The Euclidean action necessarily shows explicit time dependence}Via the identification $S_E=-2I_{os}$, the Euclidean action has a form of (\ref{Euclidean action}) and this action includes time dependent kernel $g(t)$ in it. In the usual sense of stochastic quantization, the Euclidean action is defined on the $d$-dimensional Euclidean space and stochastic process is defined in $d+1$-dimensional space since it is a product space of the real line of the stochastic time $t$ and the $d$-dimensional Euclidean space where the $S_E$ is defined on. 

The
kernel in the Euclidean action should not have the stochastic time dependence in the following sense. The purpose of the stochastic quantization is that one gets $n$-point correlation functions(for our case, 2-point correlation function only) as a consequence of the quantization of $S_E$, in the very late time of $t$. The correlation function has no  notion of stochastic time $t$ dependence since it is the correlator of the $d$-dimensional theory. The information about the stochastic time $t$ is completely washed out by taking $t\rightarrow\infty$, where the correlators are evaluated. Therefore, it may be nonsense if there is $t$ dependence in the Euclidean action, $S_E$.

By the way, the on-shell action, $I_{os}$ is in fact divergent since we did not perform holographic renormalization to obtain it  
According to the prescription in \cite{Jatkar:2013uga}, the way of getting $I_{os}$ is by computing boundary contribution of the bulk action upto the bulk equation of motion without holographic renormalization. Therefore, the on-shell action $I_{os}$ is ill-defined on the AdS boundary. 

One can resolve this issue if one defines the on-shell action on the $r=\epsilon$
hyper surface near AdS boundary where $\epsilon\ll1$. Once the on-shell action is defined on the hyper surface away from AdS boundary, it contains explicit $\epsilon$ dependence and it is promoted to stochastic time by the identification `$r=t$' when we evaluate Langevin equation or Fokker-Planck action in the stochastic frame as addressed in \cite{Oh:2012bx}. Due to this rule, it is impossible to avoid for $S_E$ not to have time dependent kernel in it.

\paragraph{Resolution of the time dependent kernel issue}
One way to resolve this issue is that one develop a mapping from the Langevin equation with explicit time dependence to that with a standard kernel without time dependence. 
Our Langevin equation is (\ref{langevin equation with g}) and the Fokker-Planck action becomes a form of harmonic oscillator with time dependence frequency $\omega(t)$, which is given by
\begin{equation}
\omega^2(t)=g^2(t)-\partial_t g(t).
\end{equation}

Now, we develop a mapping from this frame to a new stochastic frame with standard kernel without time dependence. Let us consider the following map:
\begin{equation}
\label{map to a new frame}
\Phi_k(t)\equiv u(T)\Psi_k(T), {\ }\eta_k(t)=\frac{\zeta_k(T)}{u(T)}, {\rm \ and\ }t=\int u^2(T)dT,
\end{equation}
where $\Psi_k(T)$ is a stochastic field, $\zeta_k(T)$ is a white Gaussian noise and $T$ is a stochastic time in the new stochastic frame. For the transform to the right stochastic frame, we request that $u(T)$ satisfies
\begin{equation}
g(T)=\frac{1}{u^2(T)}-\frac{\dot u(T)}{u^3(T)} {\rm \ or\ equivalently,\ } g(t)=\frac{1}{u^2(t)}-\frac{\dot u(t)}{u(t)}.
\end{equation} 
The `$\cdot$' denotes derivative with respect to its argument.
Under such a mapping, the Langevin equation(\ref{langevin equation with g}) and the 2-point correlation of white Gaussian noise(\ref{etaeta}) transform to
\begin{equation}
\label{transform result}
\frac{\partial \Psi_k(T)}{\partial T}=-\Psi_k(T)+\zeta_k(T) {\rm \ and \ }
\langle \zeta_k(T) \zeta_{k^\prime}(T^\prime)\rangle=\delta(T-T^\prime)\delta^k(k-k^\prime),
\end{equation} 
respectively. This means that in this new frame, the Langevin equation is derived
from a new Euclidean action with a kernel of identity as
\begin{equation}
S_E=\int \Psi_k\Psi_{-k}d^dk.
\end{equation}

As the simplest example, one choose 
\begin{equation}
u(t)=\frac{1}{\sqrt{\nu}}t^{1/2}, {\rm \ or \ equivalently\ } u(T)=\frac{1}{\sqrt{\nu}}e^{\frac{T}{2\nu}}
\end{equation}
for the zero boundary momentum case discussed in the previous section. Then, one can get the same answer with (\ref{transform result})
\footnote{We have omitted $\delta^d(k-k^\prime)$ in the result since it is the case
that the boundary momentum is turned off.} via the mapping (\ref{map to a new frame}) from (\ref{correlation with p=0}) and (\ref{langevin equation with p=0}).

\section{Alternative quantization and the related stochastic business}
\label{Alternative quantization and the related stochastic business}
As we addressed in the Introduction, when the scalar field mass is in the range of $-\frac{d^2}{4}\leq m^2 \leq-\frac{d^2}{4}+1$, one can have alternative quantization scheme in the dual CFTs defined on the AdS boundary. In this section, we argue that even in such a case, SQ can capture the radial evolution of the single trace operator as well as the double trace deformation correctly.

It is not surprised that to deal with SQ in the case that alternative quantization is allowed in the holographic side, there is no change except switching the Dirichlet boundary condition to other possible boundary conditions at the initial time($t=t_0$). As addressed in many of the literatures\cite{Jatkar:2012mm,Witten:2001ua,Sever:2002fk}, when the alternative quantization is possible, various boundary conditions can be imposed as a generalization of the Neumann boundary condition. 

Suppose that $\phi$ is the stochastic field. The boundary conditions that we request for SQ in the case of the standard quantization in the holographic side are
\begin{itemize}
\item At $t=t_f$, where $t_f$ is the final time in the Langevin process, we request that \\ 
$\phi(t=t_f)=\phi(t_f)$
\item At $t=t_0$, where $t_0$ is the initial time in the process, we impose that $\phi(t=t_0)=0$
\end{itemize}
This is a Dirichlet boundary condition at the initial time. For the alternative quantization, it will be shown that 
if one impose Neumann boundary condition at the initial time, the radial evolution of the boundary effective action is completely captured by SQ. It is rather natural because one of the suggested relation between SQ and HWRG in \cite{Oh:2012bx} is the identification that $t=r$ and the initial stochastic time $t_0$ corresponds to the radial location of the hypersurface near AdS boundary. Therefore imposing Neumann boundary condition to the stochastic fields at the initial stochastic time corresponds to imposing that to the field in holography on the $r=\epsilon$ hypersurface near the AdS boundary. 

\paragraph{Massless scalar in AdS$_2$}To be more precise, we revisit the case with massless scalar field in AdS$_2$ which is discussed in \cite{Oh:2012bx} for the standard quantization 
and demonstrate a process to obtain the radial evolution of the boundary single trace operator $J$ and the double trace deformation term $D$ for the alternative quantization case. The action of the massless scalar in AdS$_2$ is given by
\begin{equation}
\label{scalar-bulk-action}
S_{bulk}=\frac{1}{2}\int dr d\tau \sqrt{g} g^{\mu\nu}\partial_\mu \phi \partial_\nu \phi,
\end{equation}
where the metric $g_{\mu\nu}$ is AdS$_2$ metric, which is given by
\begin{equation}
ds^2=\frac{dr^2+d\tau^2}{r^2}.
\end{equation}
Upto the bulk equation of motion(by using its regular solutions), the on-shell action can be obtained as
\begin{equation}
I_{os}=-\frac{1}{2}\int d\omega |\omega|\phi^{(0)}_\omega\phi^{(0)}_{-\omega},
\end{equation}
which is computed in the frequency, $\omega$ space by using Fourier transform. For the alternative quantization case,
to impose Neumann boundary condition, we deform the AdS boundary by adding the following term:
\begin{equation}
S_{d}=\int_{r=\epsilon}d\omega \phi^{(0)}_{\omega}J_{-\omega}.
\end{equation}
The boundary condition is obtained by variation of the boundary action to be vanished as
\begin{equation}
\frac{\delta (I_{os}+S_d)}{\delta \phi^{(0)}_\omega}=-|\omega|\phi^{(0)}_{-\omega}+J_{-\omega}=0.
\end{equation}
Therefore, for the case of alternative quantization, rather than Dirichlet boundary condition
\begin{itemize}
\item At $t=t_0$, where $t_0$ is the initial time of the process, we can impose that 
\begin{equation}
\label{N-b}
\phi_\omega(t=t_0)=\frac{J_\omega}{|\omega|},
\end{equation}
\end{itemize}
and the initial time, $t_0$ may correspond to $r=\epsilon$ which is the location of the AdS boundary.

Except this boundary condition at the initial time, the other prescriptions for the SQ process suggested in \cite{Oh:2012bx} do not change.
Utilize the identification $S_E=-2I_{os}$ to evaluate the Euclidean action being able to derive Fokker-Planck action, find its equation of motion and the solution of it.
The solution is given by
\begin{equation}
\label{ads2-sol}
\phi^{(0)}_{-\omega}(t^\prime)=\frac{\cosh(|\omega|t^\prime)+\sinh(|\omega|t^\prime)}{\cosh(|\omega|t)+\sinh(|\omega|t)}\phi^{(0)}_{-\omega}(t),
\end{equation}
which satisfies the boundary condition at the final time, $t$ as $\phi^{(0)}_\omega(t^\prime=t)= \phi^{(0)}_\omega(t)$.
Evaluate the boundary effective action by using 
the prescription(\ref{brief-main}). Then, we have the following result:
\begin{equation}
S_B=S_{DT}+S_{J},
\end{equation}
where the $S_{DT}$ is the double trace operator part, which is given by
\begin{equation}
S_{DT}=\frac{1}{2}\int d\omega |\omega|\phi^{(0)}_\omega(t) \phi^{(0)}_{-\omega}(t)\left(\frac{\sinh(|\omega|t)+a_\omega\cosh(|\omega|t)}{\cosh(|\omega|t)+a_\omega\sinh(|\omega|t)}\right).
\end{equation}
The $S_J$ is a piece depending on the initial stochastic time $t_0$ and we manipulate this by imposing the Neumann boundary condition(\ref{N-b}). Then,
\begin{eqnarray}
S_J&=&\frac{1}{2}\int d\omega |\omega|\frac{\sinh(|\omega|t_0)+a_\omega\cosh(|\omega|t_0)}{\cosh(|\omega|t)+a_\omega\sinh(|\omega|t)}\phi^{(0)}_\omega(t)  \frac{\cosh(|\omega|t_0)+a_\omega\sinh(|\omega|t_0)}{\cosh(|\omega|t)+a_\omega\sinh(|\omega|t)}\phi^{(0)}_{-\omega}(t) \\ \nonumber
&=&\frac{1}{2}\int d\omega |\omega|\frac{\sinh(|\omega|t_0)+a_\omega\cosh(|\omega|t_0)}{\cosh(|\omega|t)+a_\omega\sinh(|\omega|t)}\phi^{(0)}_\omega(t) \phi^{(0)}_{-\omega}(t_0) \\ \nonumber
&=&\frac{1}{2}\int d\omega \frac{\sinh(|\omega|t_0)+a_\omega\cosh(|\omega|t_0)}{\cosh(|\omega|t)+a_\omega\sinh(|\omega|t)}\phi^{(0)}_\omega(t) J_{-\omega} \\ \nonumber
&\equiv&\frac{1}{2\pi}\int d\omega \phi^{(0)}_\omega(t)J_{-\omega}(t),
\end{eqnarray}
where the second equality can be understood by looking at the solution(\ref{ads2-sol}) and we impose the 
initial boundary condition(\ref{N-b}), explaining the third equality. For the last equality, we define a new quantity $J_{-\omega}(t)$ as
\begin{equation}
J_{-\omega}(t)\equiv-\frac{\tilde\beta_\omega}{\cosh(|\omega|t)+a_\omega\sinh(|\omega|t)},
\end{equation}
where $\tilde\beta_\omega$ is a constant, which is given by
\begin{equation}
\tilde\beta_\omega\equiv\pi(\sinh(|\omega|t_0)+a_\omega\cosh(|\omega|t_0))J_{-\omega}.
\end{equation}
Once we identify $\tilde\beta_\omega$ with $\beta_\omega$ given in (3.26) in \cite{Oh:2012bx}, $S_J$ precisely describe the radial evolution of the single trace part of the boundary effective action in this case.

\paragraph{Scalar field in AdS$_{d+1}$ with mass $m$ in the range that $-\frac{d^2}{4}\leq m^2 \leq-\frac{d^2}{4}+1$}
One can generalize the previous result with massless scalar field in AdS$_2$ to the massive scalar in AdS$_{d+1}$ easily. For this case, by using the technique of holographic renormalization\cite{Skenderis:2002wp,deHaro:2000xn}, one can evaluate holographic one point function which is finite. 

The regular solution of the scalar field shows a near boundary(small $r$) expansion as
\begin{equation}
\phi_p(r)=\phi_p^{(0)}r^{\frac{d}{2}-\nu}+... +\phi^{(2\nu)}_pr^{\frac{d}{2}+\nu}+...,
\end{equation}
where $\phi^{(0)}_p$ is the coefficient of non normalizable mode and $\phi^{(2\nu)_p}$ is the coefficient of the normalizable mode. In Euclidean case, they are related and in fact
\begin{equation}
\phi_p^{(2\nu)}=2^{2\nu}\frac{\Gamma(-\nu)}{\Gamma(\nu)}|p|^{2\nu}\phi^{(0)}_p,
\end{equation}
where we restrict ourselves in the case that $\nu$ is not an integral number. By adding appropriate counter terms, one can evaluate finite holographic one point function, which is given by
\begin{equation}
\langle O_p\rangle\equiv-\frac{1}{\sqrt{g}}\frac{\delta S}{\delta\phi^{(0)}_p}=2\nu 2^{-2\nu}\frac{\Gamma(-\nu)}{\Gamma(\nu)}|p|^{2\nu}\phi^{(0)}_p,
\end{equation}
where $S$ is the renormalized holographic action. When alternative quantization is possible, one can add a deformation to the conformal boundary and change the boundary condition. One one adds
\begin{equation}
S_{\rm deform}=\int \sqrt{g}2\nu2^{-2\nu}\frac{\Gamma(-\nu)}{\Gamma(\nu)}|p|^{2\nu-1}J_{-p}\phi^{(0)}_p,
\end{equation}
then Neumann boundary condition can be imposed on the conformal boundary, which is given by
\begin{equation}
\label{NnN}
|p|\phi^{(0)}_p=J_{-p},
\end{equation}
which is the similar with the Neumann boundary condition of massless scalar in AdS$_2$ (\ref{N-b}). If we request (\ref{NnN}) rather than imposing the initial boundary condition (\ref{initial-ki}), then we can correctly capture the single trace operator(\ref{double-trace radial evolution non-zero Momentum}). For the correct answer, we have defined that
\begin{equation}
\beta_p=\frac{(2\pi)^d}{2}\partial_{t_0}[t^{1/2}_0(c_1 K_\nu(|p|t_0)+c_2 I_\nu(|p|t_0))]J_{-p}.
\end{equation}



\section*{Acknowledgement}
J.H.O would like to thank Dileep P. Jatkar for useful discussion. He also thanks JHEP referee for advising me to add some of the missing points worthy to discuss in this note.
He especially thank his W.J. This article is fully supported by the research fund of Hanyang University(HY-2013).

\end{document}